\documentclass[fleqn,usenatbib]{mnras}

\usepackage{caption}
\usepackage{graphicx}	
\usepackage{amsmath}	
\usepackage{xcolor}

\title[Evidence for profile changes in PSR J1713+0747 using the uGMRT]{Evidence for profile changes in PSR J1713+0747 using the uGMRT}

\author[Singha et al.]{Jaikhomba Singha$^1$,\thanks{E-mail:mjaikhomba@gmail.com}
Mayuresh P Surnis$^2$, 
Bhal Chandra Joshi$^3$,
Pratik Tarafdar$^{4}$,
\newauthor
Prerna Rana$^5$, 
Abhimanyu Susobhanan$^5$,  
Raghav Girgaonkar$^6$, 
Neel Kolhe$^7$, 
\newauthor
Nikita Agarwal$^8$,
Shantanu Desai$^{6}$,
T Prabu$^{9}$, 
Adarsh Bathula$^{10}$, 
Subhajit Dandapat$^{5}$,
\newauthor
Lankeswar Dey$^{5}$,
Shinnosuke Hisano$^{11}$,
Ryo Kato$^{12}$, 
Divyansh Kharbanda$^{6}$,
\newauthor
Tomonosuke Kikunaga$^{11}$,
Piyush Marmat$^1$,
Sai Chaitanya Susarla$^{13}$,  
Manjari Bagchi$^{4,14}$,  
\newauthor
Neelam Dhanda Batra$^{15}$, 
Arpita Choudhury$^{4}$,
A Gopakumar$^{5}$, 
Yashwant Gupta$^3$,
\newauthor
M A Krishnakumar$^{16}$,
Yogesh Maan$^3$,
P K Manoharan$^{17}$, 
K Nobleson$^{18}$, 
Arul Pandian$^{9}$, 
\newauthor
Dhruv Pathak$^{4,14}$,
Keitaro Takahashi$^{11,19,20}$\\
$^{1}$Department of Physics, Indian Institute of Technology Roorkee, Roorkee 247667, Uttarakhand, India\\
$^{2}$Jodrell Bank Centre for Astrophysics, Department of Physics and Astronomy, The University of Manchester, Manchester, M13 9PL, UK\\
$^{3}$National Centre for Radio Astrophysics, Tata Institute of Fundamental Research, Pune 411007, Maharashtra, India\\
$^{4}$The Institute of Mathematical Sciences, CIT Campus, Taramani, Chennai 600113, Tamil Nadu, India.\\
$^{5}$Department of Astronomy and Astrophysics, Tata Institute of Fundamental Research, Mumbai 400005, Maharashtra, India\\
$^{6}$Department of Physics, Indian Institute of Technology Hyderabad, Kandi, Telangana 502285, India\\
$^{7}$Department of Physics, St. Xavier's College (Autonomous), Mumbai 400001, Maharashtra, India\\
$^{8}$Department of Electronics and Communication Engineering, Manipal Institute of Technology, Manipal Academy of Higher Education, \\ Manipal 576104, Karnataka, India\\
$^{9}$Raman Research Institute, Bengaluru 560080, Karnataka, India \\
$^{10}$Department of Physics, Indian Institute of Science Education and Research Mohali, Sector 81, Punjab 140306, India\\
$^{11}$Kumamoto University, Graduate School of Science and Technology, Kumamoto, 860-8555, Japan\\
$^{12}$Osaka City University Advanced Mathematical Institute, 3-3-138, Sugimoto, Sumiyoshi-ku, Osaka, 558-8585, Japan\\
$^{13}$School of Physics, Indian Institute of Science Education and Research Thiruvananthapuram, Vithura, Thiruvananthapuram 695551, \\ Kerala, India\\
$^{14}$Homi Bhabha National Institute, Training School Complex, Anushakti Nagar, Mumbai 400094, Maharashtra, India\\
$^{15}$Department of Physics, Indian Institute of Technology Delhi, New Delhi-110016, India\\
$^{16}$Fakult{\"a}t f{\"u}r Physik, Universit{\"a}t Bielefeld, Postfach 100131, 33501 Bielefeld, Germany\\
$^{17}$Arecibo Observatory, University of Central Florida, Arecibo, PR 00612, USA\\
$^{18}$Department of Physics, BITS Pilani Hyderabad Campus, Hyderabad 500078, Telangana, India\\
$^{19}$International Research Organization for Advanced Science and Technology, Kumamoto University, 2-39-1 Kurokami, \\ Kumamoto 860-8555, Japan\\
$^{20}$National Astronomical Observatory of Japan,
2-21-1 Osawa, Mitaka, Tokyo 181-8588, Japan
}

\date{Accepted XXX. Received YYY; in original form ZZZ}

\pubyear{2021}

\begin{document}
\label{firstpage}
\pagerange{\pageref{firstpage}--\pageref{lastpage}}
\maketitle


\begin{abstract}
PSR J1713+0747 is one of the most precisely timed pulsars in the international pulsar timing array experiment. This pulsar showed an abrupt profile shape change between April 16, 2021 (MJD 59320) and April 17, 2021 (MJD 59321). In this paper, we report the results from multi-frequency observations of this pulsar carried out with the upgraded Giant Metrewave Radio Telescope (uGMRT) before and after the event. We demonstrate the profile change seen in Band 5 (1260 MHz -- 1460 MHz) and Band 3 (300 MHz -- 500 MHz). The timing analysis of this pulsar shows a disturbance accompanying this profile change followed by a recovery with a timescale of $\sim 159$ days. Our data suggest that a model with  chromatic index as a free parameter is preferred over models with combinations of achromaticity with DM bump or scattering bump. We determine the frequency dependence to be $\sim\nu^{+1.34}$.
\vspace{0.0in} 
\end{abstract}




\section{Introduction}  \label{sec:intro}
Radio pulsars generally show little variability in their average pulse profiles obtained after adding a large number of phase-aligned individual pulses.
However, there are notable exceptions, where a pulsar switches between two or more distinct average profile shapes. 
This phenomenon, first noted by \citet{Backer1970} and \citet{Lyne1971}, is known as a profile mode change and has been reported in many slow pulsars \citep{Rankin1993,Wang2007}. 
Mode changes may be accompanied by changes in the timing behaviour of the pulsar \citep{Lyne+2010}, which can potentially degrade their timing precision and accuracy.

Pulsar Timing Array \citep[PTA:][]{FosterBacker1990} experiments, such as the European Pulsar Timing Array \citep[EPTA:][]{KramerChampion2013}, the Indian Pulsar Timing Array \citep[InPTA:][]{JoshiAB+18}, the North American Nanohertz Observatory for Gravitational Waves \citep[NANOGrav:][]{McLaughlin2013}, the Parkes Pulsar Timing Array \citep[PPTA:][]{Hobbs2013}, and the International Pulsar Timing Array \citep[IPTA:][]{Hobbs2010}, rely on precision timing of millisecond pulsars (MSPs) to detect nanohertz gravitational waves (GWs). 
While common in slow pulsars, phenomena similar to mode changes have been reported in only two millisecond pulsars till date.  
A periodic change in the profile over a timescale of about 2 s was reported for the first-detected black widow pulsar, PSR B1957+20 \citep{MahajanKM+2018}. A distinct profile change was reported in PSR J1643$-$1224 around MJD 57074 by \citet{Shannon+2016}, which introduced systematics in the timing residuals of this pulsar at 10 cm and 20 cm bands. 
Notably, this pulsar is included in the PTA experiments, and such events may hamper the PTA experiments in their quest to detect nanohertz GWs. Therefore, it is crucial to look for and accurately study and model the mode changes and similar phenomena in PTA pulsars. This will allow us to explore their implications and ensure continued high precision timing.

PSR J1713+0747 is a 4.6 ms pulsar with a low-mass white dwarf companion \citep{FosterWC+1993}, and has the second lowest measured timing noise behaviour in the PTA ensemble \citep{Perera+2019}.
However, this pulsar has shown two timing events in the past, where a jump in the timing residuals was observed followed by an exponential recovery with  timescales of about 62 days in 2008,  and 25 days in  2016 \citep{LamEG+2018,Goncharov+2020}. 
Both these events were interpreted as chromatic events, suggesting a connection with the line-of-sight passing through inter-stellar medium (ISM) structures \citep{LinLL+2021}. 
No evidence was reported for a change in the average profile of this pulsar in the first event but \cite{Goncharov+2020} reported changes in the pulse profile associated with the second event.

A profile change event was recently reported in PSR J1713+0747 by \citet{Xu+2021} using the Five hundred metre Aperture Spherical Telescope (FAST), \citet{2021ATel14652} using the Canadian Hydrogen Intensity Mapping Experiment (CHIME) and \citet{2021ATel14667} using the uGMRT.
\citet{2021ATel14652} constrained the time of the event to  between MJDs 59320 and 59321.
This source is currently being actively monitored by multiple pulsar timing groups and PTAs, which are part of the IPTA using multiple telescopes.
In this work, we report the evidence of a significant change in the pulse profile of this pulsar accompanied by a change in timing behaviour. 
The details of our observations are described in Section \ref{sec:obs} followed by a brief description of our analysis and results in Section \ref{sec:result}. 
The implications of this event are discussed in Section \ref{sec:conc}.

\begin{figure*}
    \centering
    \includegraphics[scale=0.7,trim={0  0.3cm 0  0},clip]{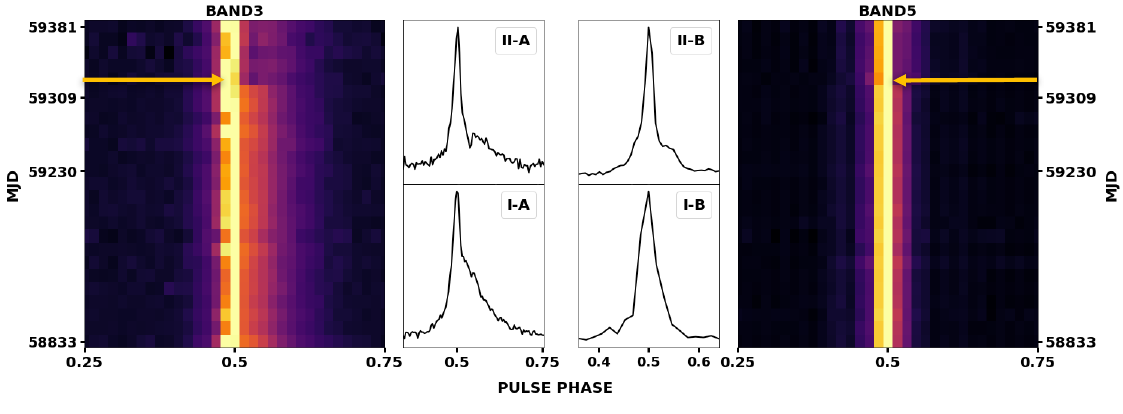}
    \caption{Profile changes observed at Band 3  and Band 5  before and after MJD 59321 are shown in this figure. The colour-map plots show the variation in profile with observation epochs indicated in MJD along the vertical axes for Band 3 (left plot) and Band 5 (right plot), respectively, where the profile change epoch is indicated by arrows.  
    The plots in the center show the frequency and time collapsed profiles before (I-A and I-B) and after (II-A and II-B) the event for Band 3 (left plots) and Band 5 (right plots) as solid lines. }
    \label{fig:modeprofs}
\end{figure*}
\section{Observations and data reduction}
\label{sec:obs}

PSR J1713+0747 was observed with the uGMRT \citep{Gupta+2017} as part of the Indian Pulsar Timing Array \cite[InPTA:][]{JoshiAB+18} observing program before and after the profile change event reported here, as well as via a Director's Discretionary Time (DDT) project after the event. 
The InPTA observations were carried out using two sub-arrays of 10 and 15 antennae at 300$-$500 MHz (Band 3) and 1260$-$1460 MHz (Band 5), respectively. The two frequency bands were covered in a single observing session simultaneously.
While the antennae in both the arrays were phased before forming a sum, Band 3 data were also coherently dedispersed to the known dispersion measure (DM) of the pulsar using a real-time pipeline \citep{DeGupta2016}. Simultaneous coverage of such widely separated frequency bands is a unique feature of the uGMRT observations.
In addition, single-band observations at Band 3 were carried out on some of the epochs allocated in the DDT project.
All observations used 200 MHz bandwidth. This bandwidth was split into 256 and 128 sub-bands (before and after MJD 59348, respectively) for Band 3 and 1024 sub-bands for Band 5 observations, respectively.

The time-series data were recorded using the GMRT Wide-band Backend \citep[GWB:][]{Reddy+2017} together with the timestamp at the start of the observation with a sampling time of 10 $\mu$s (before MJD  59348) and   5 $\mu$s (after MJD  59348) at Band 3, and 40 $\mu$s at Bands  5. 
The raw time-series were further reduced to \texttt{Timer} format \citep{vanStratenBailes2011} using the \texttt{pinta} pipeline \citep{Susobhanan+2021}. 
Further analysis was carried out using the widely-used pulsar software \texttt{PSRCHIVE} \citep{Hotan+2004}. 
The time-series were folded using pulsar ephemeris derived from the IPTA Data Release 2 \citep[IPTA DR2:][]{Perera+2019}. 
Out of the two radio-frequency mitigation methods available in \texttt{pinta}, we exclusively use \texttt{RFIClean} \citep{Maan+2020} in this work.

\section{Analysis and Results} \label{sec:result}

We used 20 observing epochs before the event with a cadence of around 14 days, and 6 epochs after the event with cadence ranging from 3-10 days.
After dedispersion and collapsing all the sub-integrations and sub-bands, the folded profiles were normalised with the  area under the profile and then stacked together to produce a single multi-epoch file for each of Band 3 and 5.
These stacked profiles are shown as colour-map plots in Figure \ref{fig:modeprofs}, where a distinct change is seen in both Band 3 and Band 5 before and after MJD 59321, reported to be the epoch of the event  \citep{2021ATel14652}. The reduced data were collapsed in frequency and time for  typical profiles before and after the event for both the bands and these are shown in the central panels of Figure \ref{fig:modeprofs}. 
We see that the broad central component has given way to a narrow central component with the appearance of a weaker broad trailing component in Band 3 after the epoch of transition (EoT). On the other hand, the leading component has weakened with the appearance of a trailing component in Band 5 data. 
This is similar to the behaviour of the mode changing multi-component profile pulsar like PSR J0332+5434 \citep{Lyne1971, Liu_2006, 2011chen}.

\subsection{Profile Analysis}
\begin{figure*}
    \centering
    \includegraphics[scale=0.42,trim={0.5cm, 0.52cm, 0.2cm, 0.55cm}]{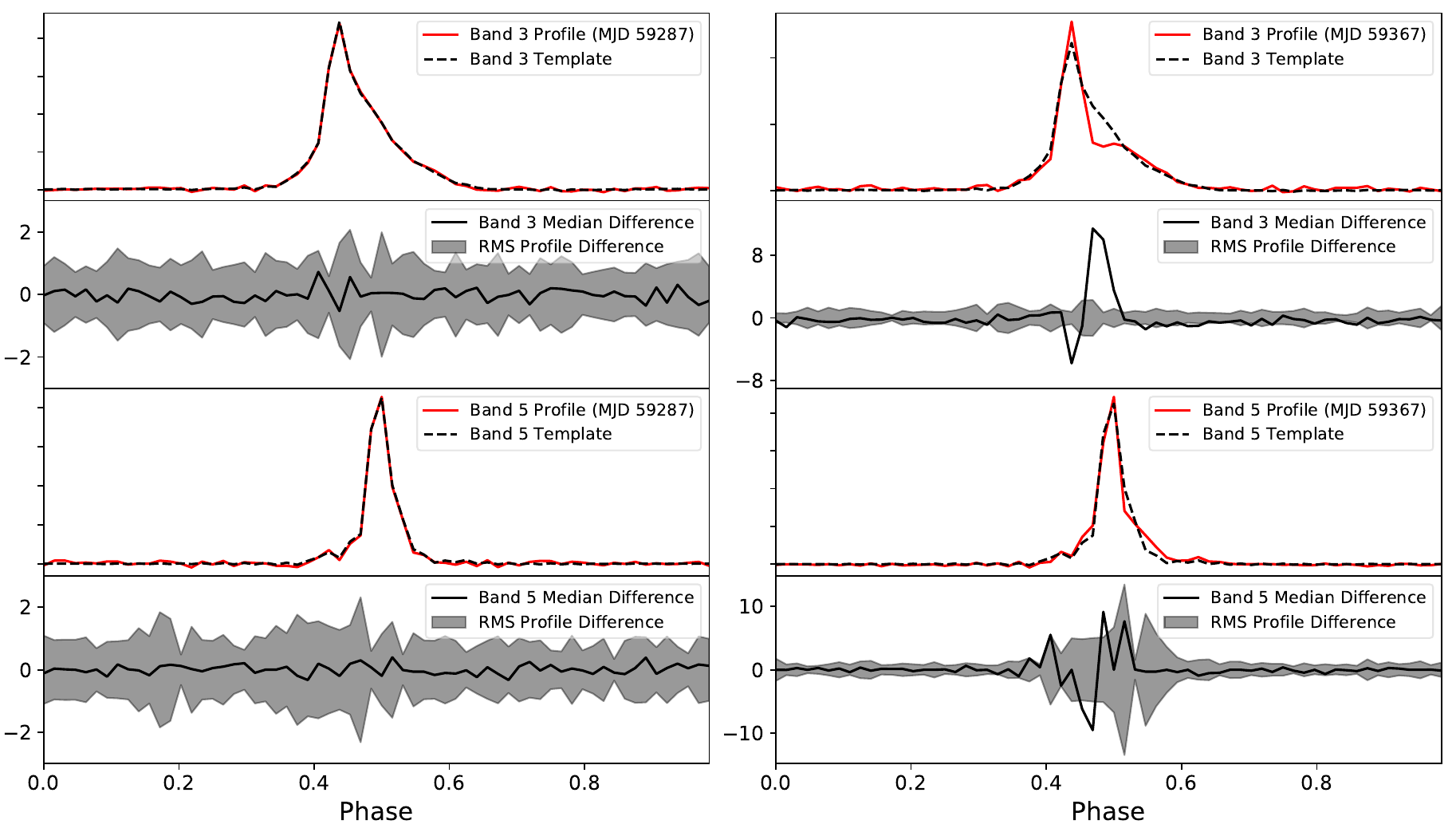}
    \caption{A comparison of profiles before (left column) and after (right column) the profile change event with templates formed from epochs before the event. 
    The first and third rows show the profile comparisons in Bands 3 and 5, respectively where we have plotted the scaled and aligned profiles from two representative epochs from before (MJD 59287) and after (MJD 59367) the event in arbitrary units.
    In the second and fourth rows, we plot respectively for Bands 3 and 5, (a) the median profile differences computed from the profiles of all epochs before and after the event normalised by the corresponding off-pulse RMS of the profiles as solid black line, and (b) the standard deviations of the profile differences normalised by the off-pulse RMS as grey filled area.
    A much higher variability ($\sim$12 times the off-pulse RMS) is seen for post-EoT profiles at Band 5 when compared to Band 3.}
    \label{fig:diffprof}
\end{figure*}
We quantified the degree of change in the profile as follows. 
First, a high signal-to-noise ratio (S/N) template profile was formed by averaging all the profiles before the event.
Then, the profiles at individual epochs, both before and after the event, were scaled and shifted to align with this template using a frequency-domain cross-correlation method \citep{Taylor1992}.
The scaled and shifted profiles were then subtracted from the template to form profile differences, and were examined for significant deviations. 
These comparisons for both the bands before and after the event are shown in Figure~\ref{fig:diffprof}.
The first and third rows in this figure show the profile for typical individual epochs before and after the EoT overlaid on the templates for Band 3 and Band 5, respectively. 
The median and standard deviation of the profile differences normalised by the off-pulse root-mean-square (RMS) obtained over all the epochs are shown in the  second and fourth rows of this figure for epochs before and after the EoT. 
A peak deviation of about 12 and 8 times off-pulse RMS is seen for the profiles after the EoT, whereas all profiles for pre-EoT epochs seem to be consistent with the template profile. 
A much higher variability ($\sim$12 times the off-pulse RMS) is evident for post-EoT profiles at Band 5 compared to Band 3. 

\subsection{Timing Analysis}
The impact of this profile change event on the timing of this \textquoteleft high-quality
timing pulsar\textquoteright{} for the PTA experiments is important
to examine. For the timing analysis, both the Band 5 and Band 3 data were collapsed
to 4 sub-bands and a single frequency-resolved template was constructed
from the highest S/N profile before the EoT in both bands. Multi-frequency times-of-arrival
(TOAs) were then obtained by cross-correlation in the Fourier domain
(Taylor 1992), and analysed using TEMPO2 (Hobbs et al. 2006; Edwards
et al. 2006) in the TCB frame with the DE436 solar system ephemeris.
The timing residuals obtained from this analysis are shown in Figure
3 as black (Band 3) and grey (Band 5) points.
The phase jump seen in the timing residuals can be attributed to the
profile change event and can be seen at all sub-bands in both
Band 3 and Band 5. A hint of recovery is also visible, suggesting
that the average profile is relaxing to its original shape. We
observe that the timing jump and the subsequent recovery are stronger
in Band 5 than in Band 3. This faster recovery following
a larger jump seen in Band 5, as compared to Band 3, is also consistent
with the higher variability of the Band 5 post-event profiles (indicated
by increased standard deviation) as seen in Figure \ref{fig:diffprof} (row 4, right
column).

We investigated the frequency dependence of the timing event
by assuming a jump plus exponential recovery models with different observing
frequency ($\nu$) dependencies of the form $\nu^{-\alpha}$. These
models can be expressed mathematically as

\[
\Delta_{\text{evt}}=A\left(\frac{1400\,\text{MHz}}{\nu}\right)^{\alpha}\exp\left[\frac{-(t-t_{0})}{\tau}\right]\Theta\left[t-t_{0}\right]\,,
\]
where $A$ is the amplitude of the event, $t_{0}$ is the EoT, $\tau$
is the recovery timescale and $\alpha$ is the chromatic index. Depending
on the sign of the amplitude $A$, the model can be a dip ($A<0$)
or a bump ($A>0$). From Figure 3, it is clear that we can immediately
rule out all pure dip models with $\alpha\ge0$, including pure achromatic
dip ($\alpha=0$), dispersion measure (DM) dip ($\alpha=2$) and scattering
dip ($\alpha=4$), since the jump is clearly stronger in Band 5 (higher
frequency) than in Band 3 (lower frequency). Therefore, we investigate
three alternative models to fit the timing jump and recovery: (A)
a model with chromatic index as a free parameter, (B) an achromatic dip + DM bump
model, and (C) an achromatic dip + scattering bump model. 
We used
the \texttt{ENTERPRISE} package \citep{Ellis+2019} to compare these models
to the timing residuals in a Bayesian framework. We assumed broad
log-uniform priors for $|A|$ and $\tau$, and a broad uniform prior for $\alpha$, whereas we used the constraint
provided by \citet{2021ATel14652} as an informative
prior for $t_{0}$. We find that, for the combined Band 3
and Band 5 data, the model A described above is preferred over model
B and model C with Bayes factors of around $403$ and $2\times10^{17}$
respectively (computed via nested sampling \citep{Feroz2009} using
the \texttt{nestle} package \citep{Barbary2021}). We also cross-checked these Bayes
factors outside of the \texttt{ENTERPRISE} framework using the \texttt{dynesty} package \citep{Speagle}, and confirmed that they agree with the \texttt{ENTERPRISE}-based
estimates. According to the Jeffreys' scale \citep{Trotta08}, these point to decisive evidence for the model with chromatic index as a free parameter (model A) providing the best description of the data, amongst the models we investigated.
The fit for model A is also plotted in Figure 3 along with
the residuals after subtracting this model. The chromatic
index in model A is estimated to be $-1.34\pm0.07$ with an amplitude
of $\sim$48 $\mu$s and a recovery timescale of $\sim$159 days.

\section{Discussion} \label{sec:conc}

We report the evidence for a profile change accompanied by a disturbance in the timing behaviour in PSR J1713+0747, which is known to be one of the three best timed pulsars
\citep{Perera+2019, Kerr+2020,Alam+2020a}. The profile change is seen in both Band 3 and Band 5 data after MJD 59321 with  a peak deviation of 12 and 8 times off-pulse RMS, and an enhanced variability in Band 5 profiles after the event. This variability appears to be linked with a jump that is followed by an exponential recovery seen in the timing residuals with a magnitude of $\sim$48 $\mu$s and a recovery timescale of $\sim$159 days.
These considerations make the present event more dramatic than the past two events seen in this pulsar.  
We find that a model with chromatic index as a free parameter suggesting a $\sim\nu^{+1.34}$  frequency dependence best describes this event, which is more consistent with its origin being at least partly magnetospheric rather than it being an ISM event. 

We infer that the event is unlikely to be due to a change in the ISM alone as 
(a) the event appears to be more pronounced in Band 5 than in Band 3 with a chromatic index of $\sim-1.34$,
and (b) a profile change is seen in both high and low frequency bands. 
Although we cannot rule out a change in pulse scatter-broadening with the current data, 
this does not appear to be the cause of the event. The situation appears to be similar to the profile mode-change event in slow pulsars like PSR J0332+5434 \citep{Lyne1971, Liu_2006, 2011chen}. 
If this is indeed the case, the tail-like feature seen before the event in the Band~3 profile is due to a profile component rather than prominent pulse broadening, implying that caution needs to be exercised in modeling ISM noise for this pulsar for PTA analysis.

Profile mode changes are believed to arise from a re-organisation of pulsar beams, probably due to global changes in the pulsar magnetosphere \citep{Timokhin2010}, and can exhibit quasi-periodicity \citep{Lyne+2010} accompanied by changes in the spin-down rate. 
This leads to higher timing noise in the pulsar
\citep{arun2017}, which has the potential to degrade the precision achievable for PTA experiments relying on the high stability of MSPs. Moreover,  \cite{Lam2021} has reported the presence of multiple profile changes following this event.
These observations motivate better modeling of these events. 

The recovery seen in the timing can probably be explained by the magnetosphere of the star relaxing to the original configuration after switching abruptly to a new state.
Such a recovery is not seen in any profile mode-changing pulsars so far. 
Hence, it provides a unique probe of magnetospheric physics, if the event can be interpreted as a profile mode change. We are continuing a high cadence monitoring campaign on the pulsar and strongly recommend continued dense monitoring in the future to establish the astrophysical reasons and implications of the present event.

\begin{figure}
    \centering
    \includegraphics[scale=0.55,trim={0.5cm 0.52cm 0.2cm 0.52cm},clip]{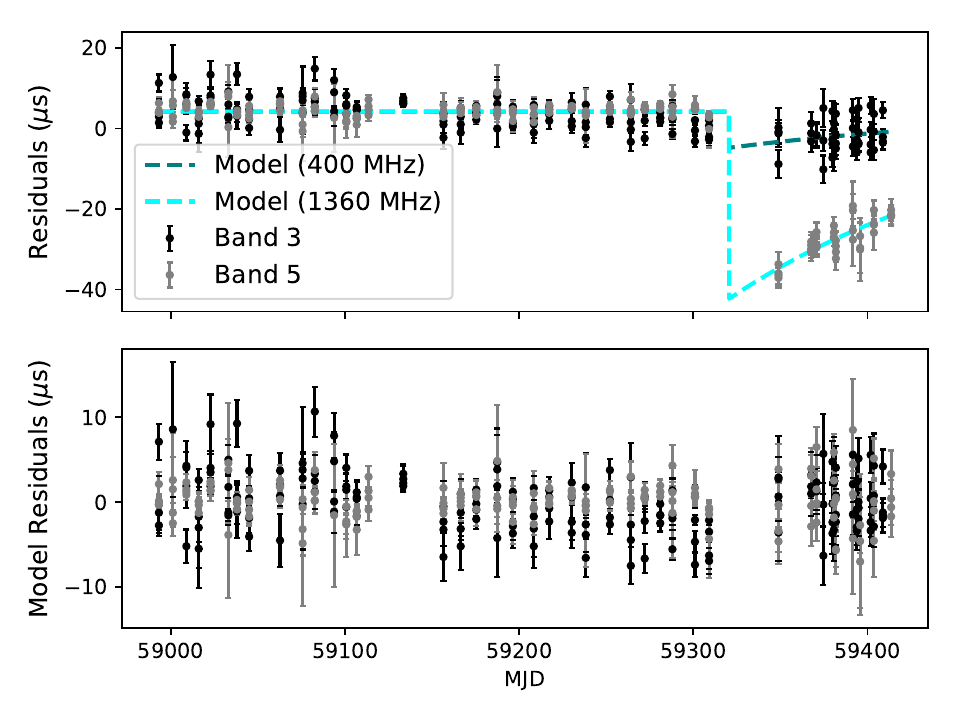}
    \caption{The timing residuals for 4 sub-bands each in Bands 3 and 5 around the profile change event are shown in this figure as black and grey points. 
    The dashed lines indicate a fit to an exponential dip model with chromatic index as a free parameter $\alpha=-1.34$. The residuals obtained by subtracting the model from the data are shown in the bottom plot. }
    \label{fig:timres}
\end{figure}

\section*{Acknowledgements}

This work is carried out by InPTA, which is part of the International
Pulsar Timing Array consortium. We thank the staff of the GMRT who made our observations
possible. GMRT is run by the National Centre for Radio Astrophysics of the Tata Institute of Fundamental Research. MPS acknowledges funding from the European Research Council (ERC) under the European Union’s Horizon 2020 research and innovation programme (grant agreement No. 694745). BCJ, PR, AS, AG, SD, LD, and YG acknowledge the support of the Department of Atomic Energy, Government  of India, under project identification \# RTI4002. BCJ and YG acknowledge support from the Department of Atomic Energy, Government of India, under project \# 12-R\&D-TFR-5.02-0700. AC acknowledge support from the Women’s Scientist scheme (WOS-A), Department of Science \& Technology, India. NDB acknowledge support from the Department of Science \& Technology, Government of India, grant SR/WOS-A/PM-1031/2014. SH is supported by JSPS KAKENHI Grant Number 20J20509. KT is partially supported by JSPS KAKENHI Grant Numbers 20H00180, 21H01130 and 21H04467, Bilateral Joint Research Projects of JSPS, and the ISM Cooperative Research Program (2021-ISMCRP-2017). We are grateful to the International Pulsar Timing Array consortium and the anonymous referee for useful feedback on the manuscript.

\section*{Data Availability}
The data underlying this article will be shared on reasonable request to the corresponding author. 




\newcommand{\ptrsl}{Phil. Trans. R. Soc. London, Ser. A}
\newcommand{\cqg}{Classical Quant. Grav.}
\bibliographystyle{mnras}
\bibliography{reference} 




\appendix


\bsp	
\label{lastpage}
\end{document}